\documentclass[aps,prd,twocolumn,groupedaddress, nofootinbib, longbibliography,superscriptaddress,amsmath,amssymb,amsfonts, floatfix]{revtex4-2}
\usepackage{graphicx} 
\usepackage[dvipsnames]{xcolor}
\usepackage{hyperref}

\definecolor{mydarkblue}{RGB}{46, 48, 146}
\hypersetup{colorlinks=true,allcolors=mydarkblue}
\usepackage{multirow}
\usepackage{adjustbox}
\usepackage[utf8]{inputenc} 
\usepackage[T1]{fontenc}    
\usepackage{microtype}      
\usepackage{tensor}
\usepackage{orcidlink}
\usepackage{float}
\usepackage[all]{hypcap}
\usepackage{subfigure}
\usepackage[capitalise]{cleveref} 
\usepackage{url}
\usepackage{orcidlink}
\usepackage{booktabs}
\usepackage{multirow}
\usepackage{aas_macros}
\usepackage{cancel}
\usepackage{ulem}

\global\long\def\DOWNCBRACE#1#2{\underset{{\scriptstyle #2}}{\underbrace{#1}}}

\begin{document}


\title{Evolution of linear matter perturbations with error-bounded bundle physics-informed neural networks}

\author{\mbox{Luca Gomez Bachar}}
\email{gomezlucajavier@gmail.com}
\affiliation{ Universidad de Buenos Aires, Facultad de Ciencias Exactas y Naturales, Departamento de Física, Av. Intendente Cantilo S/N 1428 Ciudad Autónoma de Buenos Aires, Argentina}
\affiliation{Instituto de Tecnologías en Detección y Astropartículas (CNEA, CONICET, UNSAM),
Centro Atómico Constituyentes, San Martín, Buenos Aires CP B1650KNA, Argentina}
\author{\mbox{Augusto T. Chantada}\orcidlink{0000-0002-4480-9595}}
\email{augustochantada01@gmail.com}
\affiliation{ Universidad de Buenos Aires, Facultad de Ciencias Exactas y Naturales, Departamento de Física, Av. Intendente Cantilo S/N 1428 Ciudad Autónoma de Buenos Aires, Argentina}
\affiliation{John A. Paulson School of Engineering and Applied Sciences, Harvard University, Cambridge, Massachusetts 02138, USA}
\author{\mbox{Susana J. Landau}\orcidlink{0000-0003-2645-9197}}
\affiliation{CONICET - Universidad de Buenos Aires, Instituto de Física de Buenos Aires (IFIBA), Av. Intendente Cantilo S/N 1428 Ciudad Autónoma de Buenos Aires, Argentina}
\author{\mbox{Claudia G. Scóccola}\orcidlink{0000-0002-3565-4771}}
\affiliation{Departamento de F\'isica, FCFM, Universidad de Chile, Blanco Encalada 2008, Santiago, Chile}
\author{\mbox{Pavlos Protopapas}\orcidlink{0000-0002-8178-8463}}
\affiliation{John A. Paulson School of Engineering and Applied Sciences, Harvard University, Cambridge, Massachusetts 02138, USA}

\date{\today}

\begin{abstract}
We consider the evolution of linear matter perturbations in the context of the standard cosmological model ($\Lambda$CDM) and a phenomenological modified gravity model.  We use the physics-informed neural network (PINN) bundle method, which allows to integrate differential systems as an alternative to the traditional numerical method. 
We apply the PINN bundle method to the equation that describes the matter perturbation evolution,  to compare its outcomes with recent data on structure growth, $f\sigma_8$. Unlike our previous works, we can calculate a bound on the error of this observable without using the numerical solution of the equation. For this, we use a method developed previously by ourselves to calculate an exact bound on the PINN-based solution using only the outcomes of the network and its residual. On the other hand, the use of an updated data set allows us to obtain more stringent constraints on the plane $\Omega_m-\sigma_8$ than previous works.

\end{abstract}

\maketitle

\section{Introduction}
In the past two decades, the volume and accuracy of cosmological data have grown, allowing for the restriction of the set of feasible cosmological models.
 Among these, the standard cosmological model (known as $\Lambda$CDM) is able to explain most of the current observational data with a minimum number of free parameters. However, various theoretical issues of this model have been extensively discussed in the literature, and no agreement has been reached so far. Additionally, there are tensions within the estimated values of some cosmological parameters between different data sets in the context of this model. The most discussed example is the more than 5$\sigma$ tension between the value of the Hubble constant, $H_0$, inferred from the cosmic microwave background (CMB) \cite{Planckcosmo2018} and the one obtained with type Ia supernovae and Cepheid data \cite{Riess2022}. But there are others, like the nearly $3\sigma$ tension between the value of $S_8=\sigma_8 (\frac{\Omega_m}{0.3})^{0.5}$ inferred from the CMB \cite{Planckcosmo2018} and the one obtained from weak lensing \cite{DES3year}. On the other hand, recent measurements of baryon acoustic oscillations by the DESI collaboration point out to a varying equation of state of the dark energy component of the Universe \cite{2024arXiv240403002D}. The issues highlighted here serve as the key motivation to explore alternative cosmological models, where the theoretical predictions for cosmological observables slightly differ from those of the $\Lambda$CDM model.

Concurrent advancements in both the volume and quality of cosmological data have been paralleled by significant progress in machine learning. Notably, the deployment of neural networks has considerably risen, serving as an effective instrument for evaluating data and cosmological models \cite{NN_plus_cosmo_data_exta_or_inter_1,NN_plus_cosmo_data_exta_or_inter_2,NN_plus_cosmo_data_exta_or_inter_3,NN_plus_cosmo_data_exta_or_inter_4,emulators_1}. In this paper, we focus on the development of a method that allows to integrate differential systems as an alternative to traditional numerical methods, known as physics-informed neural networks (PINNs) \cite{NN_diff_eqs,pinns, Mattheakis2019PhysicalSE}. An extension of this proposal is the so-called PINN bundle method, with which the solutions of the differential system are valid for any values of its free parameters and/or initial conditions \cite{bundlesolutions}.   We have recently applied this method to obtain the background dynamics of the Universe of five different cosmological models \cite{our_previous_paper,2024PhRvD.109l3514C}. We also showed that when the problem is computationally intense to integrate, the PINN bundle method provides a significant reduction of the computational times of the inference process. In this study, we advance by applying the method to determine the   evolution of linear matter perturbations. This marks an initial effort to address the coupled Boltzmann equations that detail the  evolution of all matter-energy constituents in the Universe. Concurrently, we employ updated observations of the quantity $f\sigma_8(z)= f(z) \sigma_8(z)$ to evaluate the predictions from two cosmological models using the PINN bundle approach.\footnote{\url{https://github.com/LucaGomez/Matter_perturbatioNNs}} We recall that $f(z)$ represents the growth rate of cosmic structure, while $\sigma_8$
is the root-mean-square linear fluctuation of the matter
distribution at the scale of $8\, \mathrm{Mpc}/h$ (where $h=\frac{H_0}{100}$ is the dimensionless
Hubble parameter). This observable  may help differentiate between the standard cosmological model and alternative gravity models, particularly at the background level, by breaking the existing degeneracy.

In our previous work, we did not calculate an error bound for the solutions obtained using the PINN bundle method. As a result, we had to rely on numerical solutions to assess the precision of our
 results, even though these numerical solutions were not used by the PINN bundle method to solve the differential system. This approach was necessary because the PINN bundle method does not yet provide an error bound for its solutions, regardless of the type of differential equation.
Error bounds have only been developed for linear ODEs, certain nonlinear ODEs, and a certain kind of first-order linear PDEs \cite{previous_work2,previous_work}. Recently, we have also developed an exact bound for a family of nonlinear first-order ODEs, of which the matter perturbation equation is a particular case \cite{2024arXiv241113848C}. In this work, we will use these results to calculate an error bound on the theoretical prediction of  $f\sigma_8(z)$ obtained using the PINN bundle method. We will show that this error bound is significantly smaller than the observational error. This allows us to use the  PINN-based solutions to perform a statistical analysis, enabling us to test the theoretical predictions using recent data of $f\sigma_8$. We will consider two theoretical models: i) the standard cosmological model and ii) a phenomenological model motivated by alternative gravity theories whose predictions can be regarded as small deviations from those of the $\Lambda$CDM model.

\textcolor{black}{In short, in our previous works \cite{our_previous_paper,2024PhRvD.109l3514C}, we demonstrated the computational advantages of the PINN bundle method, particularly its ability to significantly reduce inference times compared to traditional approaches. In this work, we built upon other previous work \cite{2024arXiv241113848C} of ours, and shift our focus to a rigorous evaluation of the method's accuracy by establishing error bounds for the predicted observables. We will thus present, for the first time as far as we know, the use self-calibrating PINN-based solutions.}

The structure of this article is as follows. In Sec.~\ref{theory}, we describe the theoretical models that we consider to apply the PINN bundle method and test with observational data. We recall the basics of the PINN bundle method and the computation of the error bound for the matter perturbation equation  in Sec.~\ref{method}. We also describe carefully the calculation of the bound on $f\sigma_8$. In Sec.~\ref{data} we describe the recent observational data we use to test our models. We discuss the results of the statistical analyses with Markov chain Monte Carlo in Sec.~\ref{results}. In the end, our conclusions are presented in Sec.~\ref{conclusions}.

\section{Theoretical models} \label{theory}

In this paper we focus on examining the evolution of linear matter perturbations. To compare our theoretical predictions with the latest observational data on the observable $f\sigma_8$, only scales significantly smaller than the cosmological horizon ($k \gg a H$) are relevant.
In this approximation, the matter perturbation evolution is decoupled from the dynamics of other matter-energy components of the Universe. We can define the linear matter density contrast as $\delta_m(t,\vec{x})=\frac{\delta\rho_m(t,\vec{x})}{\rho_{m}(t)}$, where  $\rho_{m}(t)$ is the background matter density and  $\delta\rho_m(t,\vec{x})$ represents its first order linear perturbation. It has been shown that for a generalized modified gravity model, the differential equation that describes the evolution of $\delta_m$ in Fourier space can be expressed as \cite{Tsujikawa:2007gd}:

\begin{equation}\label{fR_perturbation_equation}
    \ddot \delta_m + 2 H \dot \delta_m - 4 \pi G_{\rm eff}  \rho_m \delta_m=0
\end{equation}where the dot means the derivative respect to cosmic time, $H=\frac{\dot a}{a}$ is the Hubble parameter which is obtained by integrating the background equations of the respective cosmological model, and $G_{\rm eff}$ is the effective Newton's constant. {\color{black} Analytical solutions to Eq.\ref{fR_perturbation_equation} were obtained for specific cosmological scenarios: flat GR-based cosmologies with non-relativistic matter and radiation in \cite{silveira1994decaying} and considering only non-relativistic matter and dark energy with a constant equation of state in \cite{padmanabhan2003cosmological}.}

\subsection{Matter perturbations in \texorpdfstring{$\Lambda$CDM}{LCDM}}

In the context of the $\Lambda$CDM model, Eq.~\eqref{fR_perturbation_equation} reduces to 

\begin{equation}\label{ecdelta}
    \delta_m''(a)+h(a,\Omega_{m})\, \delta_m'(a)+g(a,\Omega_{m}) \, \delta_m(a)=0.
\end{equation}
where the primes refer to derivatives respect to the scale factor, $\Omega_m $ is the present matter density parameter, and we define 

\begin{equation}
    h(a,\Omega_{m})=a_{\rm eq}\left[\frac{2+3\frac{a}{a_{\rm eq}}+6\alpha\left(\frac{a}{a_{\rm eq}}\right)^4}{2a\left(1+\frac{a}{a_{\rm eq}}+\alpha\left(\frac{a}{a_{\rm eq}}\right)^4\right)}\right],
\end{equation}

and

\begin{equation}
    g(a,\Omega_{m})= -\left[\frac{3a_{\rm eq}}{2a\left(1+\frac{a}{a_{\rm eq}}+\alpha\left(\frac{a}{a_{\rm eq}}\right)^4\right)}\right].
\end{equation}

Also,

\begin{equation}
    a_{\rm eq}(\Omega_{m})=\frac{\Omega_{r}}{\Omega_{m}}, \quad\alpha(\Omega_{m})=a_{\rm eq}(\Omega_{m})^4 \, \frac{\Omega_{\Lambda }(\Omega_{m})}{\Omega_{r}}.
\end{equation}
%
The value of the present radiation density parameter is fixed to $\Omega_{r}=2.47 \times 10^{-5}\, h^{-2}$, in agreement with the value estimated by COBE \cite{2009ApJ...707..916F}. Furthermore, assuming a flat universe leads to a relation between the present dark energy density parameter ($\Omega_\Lambda$) and the present matter density parameter ($\Omega_{m}$) given by:  $\Omega_{\Lambda}=1-\Omega_{r}-\Omega_{m}$.

Solving Eq.~\eqref{ecdelta} yields the solution expressed as the function $\delta_m(a,\Omega_{m})$. We use this solution to perform parameter inference through the observable $f\sigma_8$, which is defined as
%
\begin{eqnarray}\label{fsigma8}
    f\sigma_8(a,\Omega_{m},\sigma_8)&=& f(a) \,\, \sigma_8(a) \nonumber \\
                                     &=& \frac{d\ln\delta(a)}{d\ln a} \,\,\sigma_8 \frac{\delta(a)}{\delta(1)}\nonumber \\
    &=& \sigma_8 \,\, a \,\, \frac{\delta_m'(a,\Omega_{m})}{\delta_m(1,\Omega_m)}.
\end{eqnarray}where $\sigma_8=\sigma_8(a=1)$.

The initial conditions are imposed during the matter-dominated era, where it holds that:

\begin{subequations}\label{ci}
\begin{align}
    \delta_m(a_{\rm ini})=a_{\rm ini},\label{ci1} \\
    \delta_m'(a_{\rm ini})=1.\label{ci2}
\end{align}
\end{subequations}
In this work, we use $a_{\rm ini}=10^{-3}$. 

The initial conditions also show a dependence with the Fourier mode $k$. However, since the matter perturbation equations in our work do not depend on $k$, we can safely assume Eq.~\eqref{ci} without affecting the prediction of the observable $f\sigma_8$.

%

\subsection{Phenomenological model of modified gravity}
\label{model_modgrav}
As mentioned in the introduction, current tensions in cosmology and the recent release of the DESI collaboration point  to a cosmological model with small deviations with respect to the $\Lambda$CDM predictions. Therefore, in this work, we assume a phenomenological model where the background has a $\Lambda$CDM behavior, while the matter perturbation equation is modified with an effective gravitational constant as proposed in Ref.~\cite{2017PhRvD..96b3542N}. In this context, Eq.~\eqref{fR_perturbation_equation} written in terms of derivatives of the scale factor reads: 

\begin{equation}\label{ecdeltaMG}
    \delta_m''(a)+h(a,\Omega_{m})\, \delta_m'(a)+g(a,\Omega_{m})\frac{G_{\rm eff}(a,g_a)}{G} \,\delta_m(a)=0,
\end{equation}where
\begin{equation}\label{geff}
    \frac{G_{\rm eff}}{G}(a,g_a) = 1 + g_a \, (1-a)^n - g_a \, (1-a)^{2n}.
\end{equation}

Here $g_a$ and $n$ are free parameters of the model. The variation in $G$ described by this last equation is similar to that of modified gravity models. Specifically, viable $f(R)$ models \cite{Hu-Sawicki,Starobinsky}  show this behavior in the subhorizon limit. Furthermore, the phenomenological model needs to satisfy the following conditions:

\begin{itemize}
\item $G_{\rm eff} > 0 $ so that the gravitons carry positive energy
\item $\dfrac{\dot G}{G}(a=1) \sim 0$ to be in agreement with solar system tests \cite{2012SoSyR..46...78P,2021Univ....7...34B}.
\item $|\dfrac{G_{\rm eff}}{G}(a) -1| < 7.36 \times 10^{-3} $ over the redshift interval $(0,1.5)$ to be in agreement with the bounds on $G_{\rm eff}$ from helioseismology \cite{1998ApJ...498..871G}.
\end{itemize}

The second requirement implies that $n\ge 2$ and therefore in this work we fix $n=2$.
Guenther et al. \cite{1998ApJ...498..871G} compared the p-mode oscillation spectra of solar models, with the
 solar p-mode frequency observations to obtain a bound the variation in G over $4.6 \times 10^{9}$ yrs. This time interval, for $\Lambda$CDM, corresponds to the redshift interval of the data considered in this work (see Section \ref{data}) and this is the reason to consider it here. Furthermore, the derivations that establish this limit are independent of any particular cosmological model.\footnote{Other works \cite{1998ApJ...498..871G} consider the nucleosynthesis bound on  ${\dot G}/{G}$ but this bound is not independent of the assumed cosmological model. 
On the other hand, the variation in $G$ is bounded for a time interval that is much longer than the one considered in this paper.} Consequently, the stringent limit on the variation of $G$ from helioseismology implies a sharp limit on the parameter $g_a$ that we will implement as an informative prior on this parameter (see section \ref{results}). 
Two final clarifications:
i) Although $f\sigma_8$ is defined in Eq.~\eqref{fsigma8}, here it also depends on the parameter $g_a$, i.e., $f\sigma_8(a) = f\sigma_8(a, \Omega_m, \sigma_8, g_a)$. ii) To solve Eq.~\eqref{ecdeltaMG}, we use the initial conditions specified for the $\Lambda$CDM model.

\section{Method}\label{method}

\subsection{Solving the differential equation with the PINN bundle method}

As in our previous works \cite{our_previous_paper, 2024PhRvD.109l3514C}, we solve the differential system describing the evolution of matter perturbations using the PINN bundle method,   an extension of physics-informed neural networks (PINNs) implemented through the NeuroDiffEq library \cite{outdated_neurodiff_ref}.

A particular feature of Eqs.~\eqref{ecdelta} and \eqref{ecdeltaMG} is that the range of integration of the independent variable $(a_{\rm ini},a_{\rm fin})=(10^{-3},1)$ involves several orders of magnitude. Also, a preliminary analysis showed that the corresponding solutions are dominated by an exponential behavior. Therefore, to facilitate the convergence of the PINN method, we introduce the following  variable transformations

\begin{equation}
    \Tilde{N} = \ln(a), \quad\quad  X = \ln(\delta_m).
\end{equation}

We also rescaled the new independent variable
\begin{equation}
    N = \frac{\Tilde{N}}{n_{\rm ini}}
\end{equation}
where $n_{\rm ini}=|\ln(10^{-3})|$, ensuring values range from $-1$ to $0$, preventing saturation of the PINNs activation functions. With these variable changes, Eq.~\eqref{ecdeltaMG} reads

\begin{equation}
    \frac{d^2X}{dN^2}+\left(\frac{dX}{dN}\right)^2+h(N,\Omega_m) \,\frac{dX}{dN}+g(N,\Omega_m) \, \frac{G_{\rm eff}(N,g_a)}{G}=0.
\end{equation}
Note that Eq.~\eqref{ecdelta} can also be expressed as the last equation with $G_{\rm eff}=G$.

It is known, and we also observed this, that solving differential equations with neural networks yields better results when they are expressed in first-order form. This approach avoids second derivatives, which can lead to complex computational graphs and increased instability. Therefore, as usual, we can rewrite the previous equation as follows:

\begin{subequations}\label{XY}
\begin{align}
    \frac{dY}{dN} + Y^2 + h(N,\Omega_m)Y + q(N,\Omega_{m},g_a) = 0,\label{ec1} \\
    \frac{dX}{dN} - Y = 0,\label{ec2}
\end{align}
\end{subequations}
where a new variable $Y$ has been introduced and $q(N,\Omega_{m},g_a)=g(N,\Omega_m)\frac{G_{\rm eff}(N,g_a)}{G}$. The initial conditions for this system are

\begin{subequations}\label{CI}
\begin{align}
    Y_0=-\ln{a_{\rm ini}},\label{CI1}\\
    X_0=\ln{a_{\rm ini}}.\label{CI2}
\end{align}
\end{subequations}

To implement the PINN bundle method, we consider two fully connected neural networks such that the output of each network corresponds to one of the two dependent variables of the differential system in Eq.~\eqref{XY}. For the $\Lambda\rm{CDM}$ model, each network has two inputs: the independent variable $N$ and the cosmological parameter $\Omega_{m}$, called the bundle parameter. For modified gravity, another bundle parameter is added, $g_a$, resulting in 3 input parameters. Each neural network has two hidden layers with 32 neurons each and a single neuron in the output layer. The PINN bundle method is framed as an optimization problem, where the task is to find the set of internal parameters of the network that minimize a cost (or loss) function $L$ which measures how effectively the PINN’s output satisfies the differential system. A common way to defining this loss function is

\begin{equation}
    L(\tilde{u},N,\boldsymbol{\theta})=\sum_i^M \mathcal{R}_i(\tilde{\boldsymbol{u}},N,\boldsymbol{\theta})^2,
\end{equation}
where $\tilde{\boldsymbol{u}}=\tilde{\boldsymbol{u}}(N,\boldsymbol{\theta})=({\tilde Y}(N,\boldsymbol{\theta}),{\tilde X}(N,\boldsymbol{\theta}))$ represents the approximate solution of the differential system.  $(\tilde{Y}, \tilde{X})$ are reparametrizations of the outputs of the networks, $(Y_\mathcal{N}, X_\mathcal{N})$, designed to satisfy the initial conditions.

\begin{subequations}\label{rep}
\begin{align}
    \tilde{Y}(N,\boldsymbol{\theta})=Y_0+(1-e^{-(N+1)}){{Y}}_{\mathcal{N}}(N,\boldsymbol{\theta}),\label{rep1} \\
    \tilde{X}(N,\boldsymbol{\theta})=X_0+(1-e^{-(N+1)}){{X}}_{\mathcal{N}}(N,\boldsymbol{\theta}).\label{rep2}
\end{align}
\end{subequations}
${\mathcal{R}}_i$ denotes the residual of the $i$-th equation, which is computed by evaluating Eqs.~\eqref{ec1} and \eqref{ec2} using the network outputs. Consequently, the right-hand side of these equations is not zero, but corresponds to the residuals.

Furthermore, $\boldsymbol{\theta}$ denotes the bundle parameters ($\boldsymbol{\theta}=\Omega_m$ in the $\Lambda \rm{CDM}$ model, and $\boldsymbol{\theta}=(\Omega_m,g_a)$ in the modified gravity model).

\subsection{Error bounds}

So far, we have described the procedure to obtain the solutions of Eq.~\eqref{XY} using the PINN bundle method. In this section, we outline the steps to obtain a bound on the error when approximating the observable $f\sigma_8$. In our previous work \cite{2024arXiv241113848C}, we provided a detailed derivation of the error bound for the dependent variable $Y$, which we denote as $\mathcal{B}_Y(N)$. Here, we briefly highlight the key steps from that derivation for reference. The error, $\eta^Y$, is defined as the difference between the true value, $Y$ and the neural network’s estimate $\tilde{Y}$. We can write $\eta^Y(N)$ as an infinite sum of unknown functions $\eta^Y_i(N)$:

\begin{equation}\label{defetai}
    \eta^Y(N)=Y-\tilde{Y}=\sum_{i=0}^\infty\eta^Y_i(N).
\end{equation}
Using this definition and Eq.~\eqref{ec1}, we can define a system of infinite differential equations for the unknown functions $\eta^Y_i(N)$, which for a particular choice of $\eta^Y_i(N)$, can be solved by direct integration. In Ref.~\cite{2024arXiv241113848C}, we showed that it is possible to define a bound as a function of $N$ and the truncation order $J_Y$:\footnote{In Ref.~\cite{2024arXiv241113848C} we demonstrate that $\eta^Y_j$ are calculated using only $\tilde Y$ and its residual.}

\begin{equation}\label{BY}
\begin{split}
    |\eta^Y(N)|\leq \mathcal{B}_Y(N;J_Y):=\left|\sum_{j=0}^{J_Y}\eta^Y_j(N)\right|+ \\
    +\frac{R\left[R(N-N_{\rm ini})\right]^{J_Y+1}}{1-R(N-N_{\rm ini})}.
\end{split}
\end{equation}
Here $N_{\rm ini}$ is the initial value of the independent variable. The value of $J_Y$ is chosen so that the second term of Eq.~\eqref{BY} remains below $10^{-8}$. This implies  $J_Y = 3$ for both cosmological  models studied in this paper. Meanwhile, $R$ is defined as

\begin{equation}
    R = \int_{-1}^0|\mathcal{R}_Y(N)|dN,
\end{equation}
where $\mathcal{R}_Y(N)$ is the residual of Eq.~\eqref{ec1}. This bound holds as long as the condition $R<1$ is satisfied.

In Appendix \ref{boundX} we show  that the bound corresponding to the variable $X$ obtained with the PINN-based method can be expressed as:

\begin{equation}\label{BX2}
\begin{split}
    \mathcal{B}_X(N;J_X):=&\left|-\int_{-1}^N\mathcal{R}_X(N^\prime)dN^\prime+\sum_{j=0}^{J_X}\int_{-1}^N\eta^Y_j(N^\prime)dN^\prime\right| \\
    &-\left[R(N+1)\right]^{J_X+1}\ln{[1-R(N+1)]}
\end{split}
\end{equation}
where $\mathcal{R}_X$ is the residual of Eq.~\eqref{ec2}. Using our trained networks the first term in Eq.~\eqref{BX2} is of order $10^{-3}$. Therefore,  we chose $J_X=3$  so that the second term in Eq.~\eqref{BX2} remains below $10^{-8}$.

To obtain the bound on the error of the observable quantity $f\sigma_8$ calculated from the outputs of the PINN bundle method, we first write: 
\begin{eqnarray}\label{fs8XY}
     f\sigma_8(a)=\sigma_8 \frac{Y(a)}{n_{\rm ini}}e^{\left(X(a)-X(a=1)\right)}.
\end{eqnarray}Therefore, we can consider $f\sigma_8$ as a function that depends on three variables: $X,U,Y$; we define here $U(\boldsymbol{\theta})=X(a=1,\boldsymbol{\theta})$ . To calculate the bound we use the mean value theorem for a scalar function that takes multiple variables. Let $f : \mathbb{R}^d \rightarrow \mathbb{R}$. We consider $\boldsymbol{p}$ and  $\boldsymbol{q}$ $\in \mathbb{R}^d$ and denote $L$  the line
segment in $\mathbb{R}^d$ with $\boldsymbol{p}$ and $\boldsymbol{q}$ as endpoints. If we assume that $f$ is continuous on $L$ and that its first derivatives are defined on $L$ (except possibly at its endpoints), then there exists a point $\boldsymbol{r} \in L$ so that:
\begin{equation}
    f(\boldsymbol{q})-f(\boldsymbol{p})=||\vec{\nabla}f(\boldsymbol{r})||(\boldsymbol{q}-\boldsymbol{p}).
\end{equation}
Then, taking $\boldsymbol{q}=(X,U,Y)$ and $\boldsymbol{p}=(\hat{X},\hat{U}, \hat{Y})$ we have:

\begin{equation}
\begin{split}\label{boundfsigma8}
      |\eta_{f\sigma_8}(a,\boldsymbol{\theta})| =& \sigma_8 |\widetilde{f\sigma_8}(X,U,Y) - \widetilde{f\sigma_8}(\hat X,\hat{U},\hat Y)| \\
      =& \sigma_8 \,\,||\vec\nabla \widetilde{f\sigma_8}(\boldsymbol{r})||\,\, ||(\eta_X(a,\boldsymbol{\theta}), \eta_U(\boldsymbol{\theta}),\eta_Y(a,\boldsymbol{\theta}))||  \\
      \leq& \,\,\sigma_8 \,\,{ \max_{\boldsymbol{s} \in D}} ||\vec \nabla \widetilde{ f\sigma_8}(\boldsymbol{s})||\\
      &\times||(\mathcal{B}_X(a,\boldsymbol{\theta}),\mathcal{B}_U(\boldsymbol{\theta}),\mathcal{B}_Y(a,\boldsymbol{\theta}))||.   
\end{split}
\end{equation}

This last expression, that bounds $|\eta_{f\sigma_8}(a,\boldsymbol{\theta})|$, is the one we are going to use in this work, and we notate it as $\mathcal{B}_{f\sigma_8}(a, \boldsymbol{\theta})$. Here $\widetilde{f\sigma_8}(a)=f\sigma_8(a)/\sigma_8$ and $D$ is an arbitrary domain that contains $L$. Ideally, one would like to use $D = L$, but because we do not know $X$, $U$ and $Y$, this is not possible. Nevertheless, we can use the bounds we have computed before for the error of $\hat{X}$, $\hat{U}$ and $\hat{Y}$ to construct a region that contains $L$. Given that $|X(a,\boldsymbol{\theta})-\hat{X}(a,\boldsymbol{\theta})|=|\eta^X(a,\boldsymbol{\theta})|\leq \mathcal{B}_X(a,\boldsymbol{\theta})$, then $X(a,\boldsymbol{\theta})\in L_X(a,\boldsymbol{\theta}):=[\hat{X}(a,\boldsymbol{\theta})-\mathcal{B}_X(a,\boldsymbol{\theta}), \hat{X}(a,\boldsymbol{\theta})+\mathcal{B}_X(a,\boldsymbol{\theta})]$. Analogously, we can define $L_U(\boldsymbol{\theta}):=[\hat{U}(\boldsymbol{\theta})-\mathcal{B}_U(\boldsymbol{\theta}), \hat{U}(\boldsymbol{\theta})+\mathcal{B}_U(\boldsymbol{\theta})]$ and $L_Y(a,\boldsymbol{\theta}):=[\hat{Y}(a,\boldsymbol{\theta})-\mathcal{B}_Y(a,\boldsymbol{\theta}), \hat{Y}(a,\boldsymbol{\theta})+\mathcal{B}_Y(a,\boldsymbol{\theta})]$ that are guarantied to contain $U(\boldsymbol{\theta})$ and $Y(a, \boldsymbol{\theta})$, respectively. With this in mind, we can now define the region $S(a,\boldsymbol{\theta}):=L_X(a,\boldsymbol{\theta})\times L_U(\boldsymbol{\theta})\times L_Y(a,\boldsymbol{\theta})$ and thus set $D=S$. It can be shown that $L \subseteq S$.

\begin{figure*}
    \centering
    \subfigure[]{\includegraphics[width=0.49\linewidth]{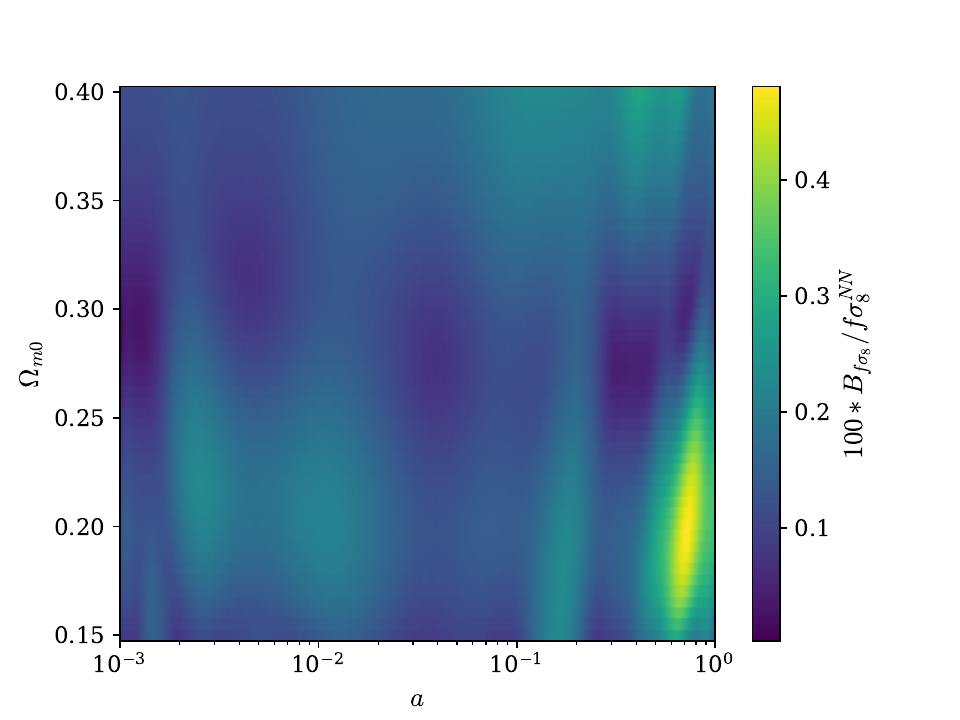}}
    \subfigure[]{\includegraphics[width=0.49\linewidth]{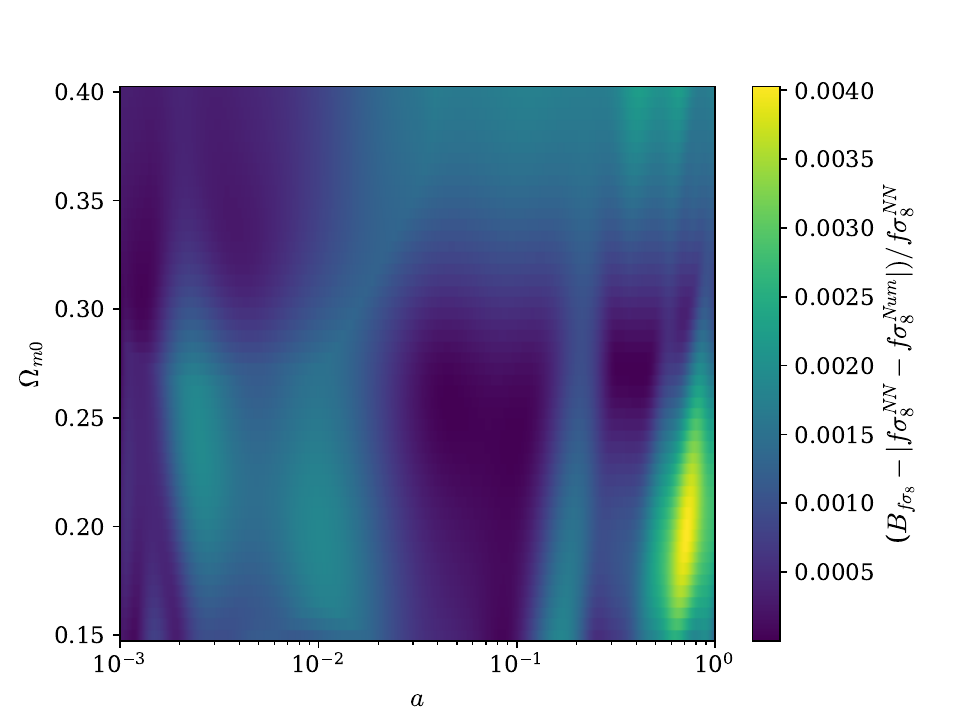}}
    \caption{Estimation of the error on the PINN approximation for the $\Lambda$CDM model as a function of the scale factor  $a$ and $\Omega_m$. Left: $100\frac{\mathcal{B}_{f\sigma_8}}{f\sigma_8}$ Right:  Difference between $\frac{\mathcal{B}_{f\sigma_8}}{f\sigma_8}$ and the relative difference  the PINN-based solution  and the ground truth $|{\frac{f\sigma_8^{\rm NN}-f\sigma_8^{\rm Num}}{f\sigma_8^{\rm NN}}}|$).
    }
    \label{BoundLCDM}
\end{figure*}

Figure \ref{BoundLCDM}  shows $100\frac{\mathcal{B}_{f\sigma_8}}{f\sigma_8}$ using the bound defined in Eq.~\eqref{boundfsigma8} together with the difference between $\frac{\mathcal{B}_{f\sigma_8}}{f\sigma_8}$ and the relative difference between the numerical solution and the outcome of the PINN bundle method 
$|\frac{f\sigma_8^{\rm NN} - f\sigma_8^{\rm NUM}}{f\sigma_8^{NN}}|$ for the $\Lambda$CDM model. This last quantity can be regarded as an estimation of $\frac{\eta_{f\sigma_8}}{f\sigma_8}$ provided that we consider the numerical solution as the ground truth.

\begin{figure*}
    \centering
    \subfigure[]{\includegraphics[width=0.49\linewidth]{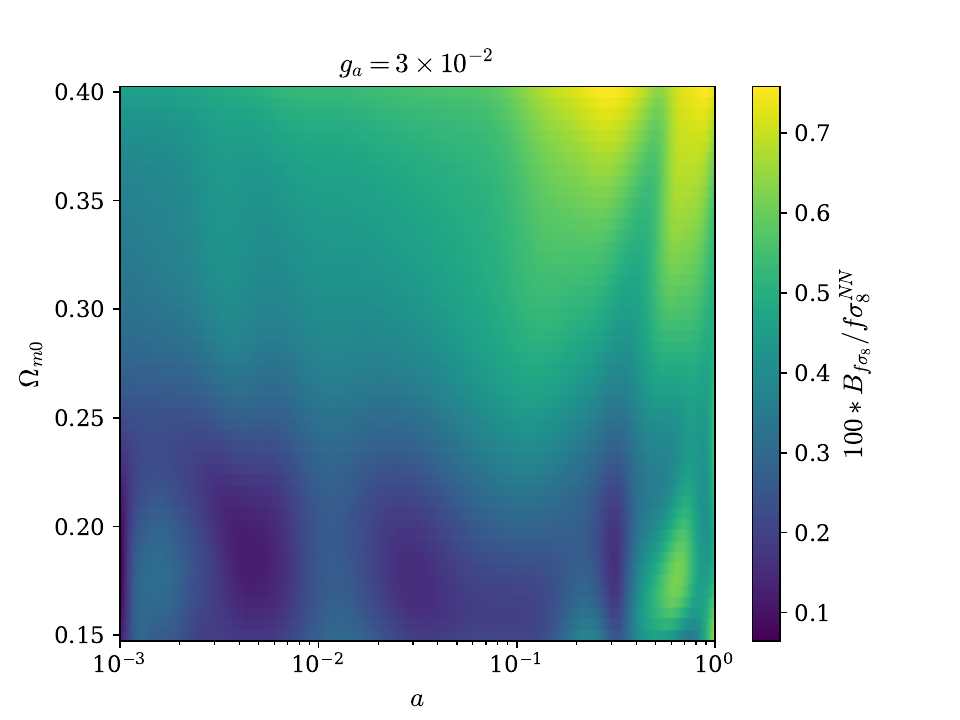}}
     \subfigure[]{\includegraphics[width=0.49\linewidth]{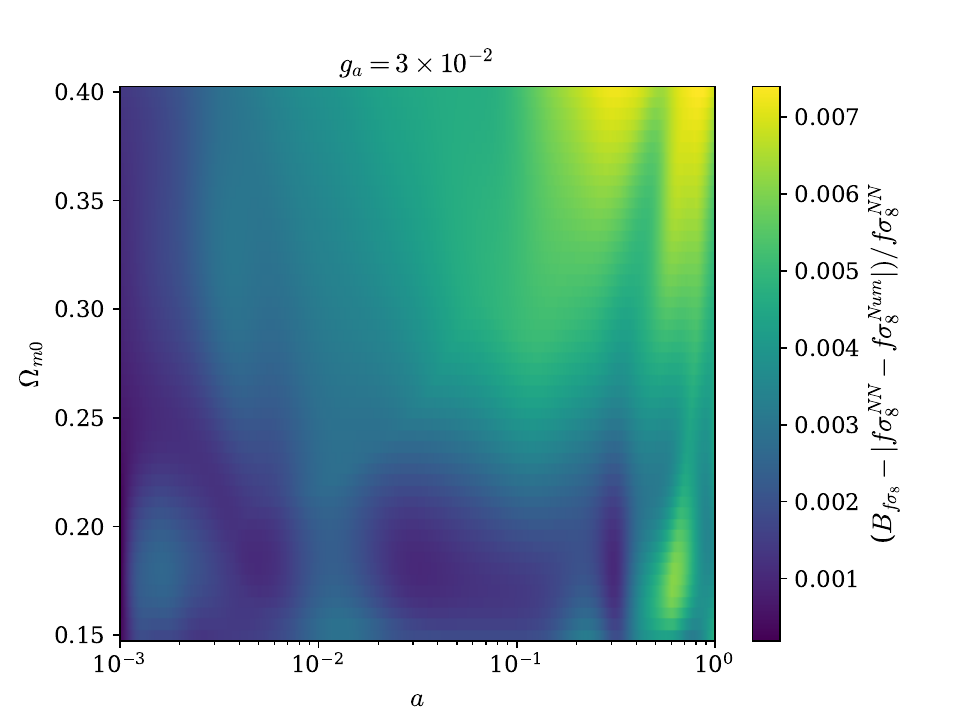}}
    \caption{Estimation of the error on the PINN approximation for the modified gravity model as a function of the scale factor $a$ and $\Omega_m$. Left: $100\frac{\mathcal{B}_{f\sigma_8}}{f\sigma_8}$ Right: Right: Difference between $\frac{\mathcal{B}_{f\sigma_8}}{f\sigma_8}$ and the relative difference  the PINN-based solution  and the ground truth $|{\frac{f\sigma_8^{\rm NN}-f\sigma_8^{\rm Num}}{f\sigma_8^{\rm NN}}}|$). Here we fix $g_a=0.03$. 
    }
    \label{BoundMGpos}
\end{figure*}

\begin{figure*}
    \centering
    \subfigure[]{\includegraphics[width=0.49\linewidth]{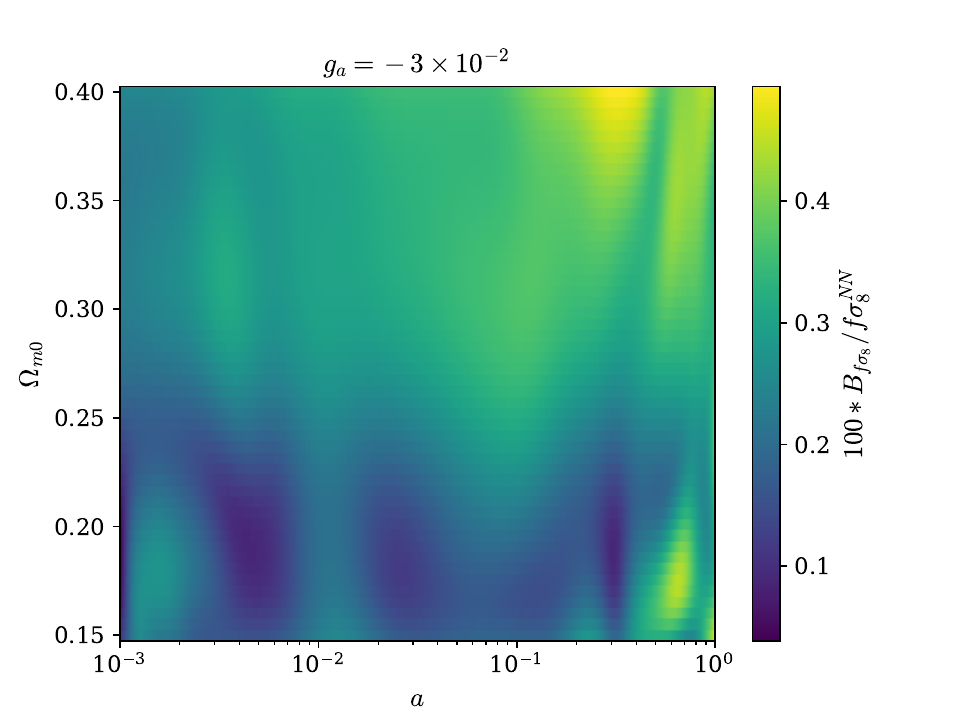}}
    \subfigure[]{\includegraphics[width=0.49\linewidth]{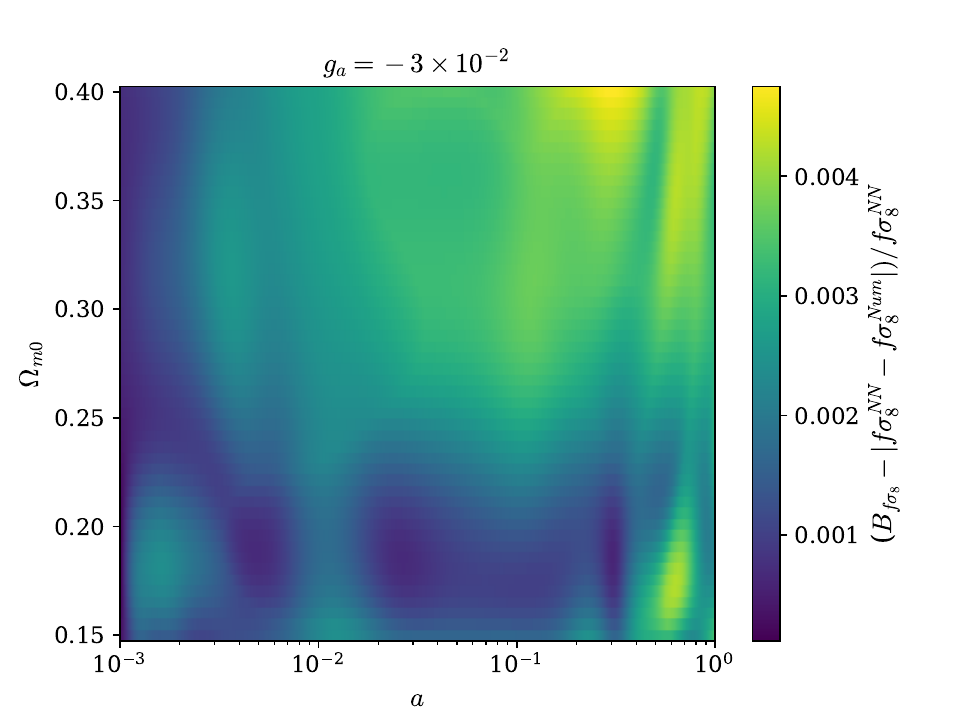}}
    \caption{Estimation of the error on the PINN approximation for the modified gravity model as a function of the scale factor $a$ and $\Omega_m$. Left: $100\frac{\mathcal{B}_{f\sigma_8}}{f\sigma_8}$ Right:  Difference between $\frac{\mathcal{B}_{f\sigma_8}}{f\sigma_8}$ and the relative difference  the PINN-based solution  and the ground truth $|{\frac{f\sigma_8^{\rm NN}-f\sigma_8^{\rm Num}}{f\sigma_8^{\rm NN}}}|$). Here we fix $g_a=-0.03$.}
    \label{BoundMGneg}
\end{figure*}

Figs.~\ref{BoundMGpos} and \ref{BoundMGneg} show the same quantities for the phenomenological modified gravity model, where we fix $g_a$ to $0.03$ and $-0.03$.
Note that $\mathcal{B}_{f\sigma_8} \propto \sigma_8$ and $f\sigma_8 \propto \sigma_8$. In addition, the error on $f\sigma_8$ is derived from the errors of the networks, and it is important to note that Eqs.~\eqref{XY} are independent of $\sigma_8$.
Therefore, neither $\frac{\mathcal{B}f\sigma_8}{f\sigma_8}$ nor the relative difference with the numerical solution depends on it.

As shown in Figures \ref{BoundLCDM}, \ref{BoundMGpos}, and \ref{BoundMGneg} the estimation computed here of the percentage error in $f\sigma_8$, namely $100\frac{\mathcal{B}_{f\sigma_8}}{f\sigma_8}$  is below $0.4\%$ for  $\Lambda$CDM and under $0.7\%$  for the modified gravity model. For modified gravity, we verified this for the entire range of $g_a$ considered in the statistical analysis in Sec.~\ref{results}.
Additionally, we verified that for all the data used in this paper (see Table \ref{tab:data-fs8} in Sec.~\ref{data}), the following condition holds
\begin{equation}\label{cond1}
    \mathcal{B}_{f\sigma_8}(z_i,\boldsymbol{\theta},\sigma_8) \ll {\min_{i}{ [\sigma_i]}} \quad\forall z_i \wedge \forall  (\boldsymbol{\theta}, \sigma_8) \in \Pi.
\end{equation}
In this context, $\min_{i}{ [\sigma_i]}$ denotes the smallest value among the dataset presented in Table \ref{tab:data-fs8}, while $\Pi$ is the domain encompassing the cosmological parameters. Consequently, under these conditions, we can safely use our PINN-based solutions in the statistical analysis without considering its uncertainties. 
However, if the condition Eq.~\eqref{cond1} is not satisfied, it would be possible to use a less restrictive condition, yet still valid.

\begin{equation}\label{cond2}
    \mathcal{B}_{f\sigma_8}(z_i,\boldsymbol{\theta},\sigma_8) < \varepsilon\sigma_i \quad\forall i \wedge \forall  (\boldsymbol{\theta}, \sigma_8) \in \Pi,
\end{equation}
with $\varepsilon\ll 1$. This condition could be applied after the training of the networks to ensure that the errors on the PINN-based solution are sufficiently small or it could be used during the training. In this latter case, one could define a value of $\varepsilon$ and use Eq.~\eqref{cond2} as a stopping criterion for the network training.

\section{Data}\label{data}

The growth rate of cosmic structure, parameterized as $f\sigma_8$, is a key observable in cosmology, providing insights into the dynamics of large-scale structure formation and the underlying matter content of the Universe. Here, $f$ represents the linear growth rate of matter perturbations, which depends on the rate of expansion and the nature of gravity, while $\sigma_8$ quantifies the amplitude of matter density fluctuations on scales of $8 h^{-1} \rm Mpc$. The combination $f\sigma_8$ is particularly useful, as it directly relates to the velocity field of matter and is largely independent of galaxy bias, making it a robust probe for testing gravity models and constraining cosmological parameters. In Table~\ref{tab:data-fs8}, we present a compilation of $f\sigma_8$  measurements from various observations, which will be used to compare theoretical predictions with observational data.

\begin{table}[!h]
\begin{center}
\setlength{\tabcolsep}{4pt}
\renewcommand{\arraystretch}{1.35}
\caption{Compilation of currently available growth rate data. \label{tab:data-fs8}}
\begin{ruledtabular}
\begin{tabular}{lcccc}
$z$ & $f \sigma_{8} \pm \sigma_{f \sigma_{8}}$ & Fiducial $\Omega_\mathrm{m}$ & Ref. \\ 
\hline   
$0.17$ & $0.510 \pm 0.060$ & $0.3$ & \cite{Song:2008qt} \\
$0.02$ & $0.314 \pm 0.048$ & $0.266$ & \cite{Davis:2010sw} \\
  $0.02$ & $0.398 \pm 0.065$ & $0.3$ & \cite{Turnbull:2011ty} \\
  $0.44$ & $0.413 \pm 0.080$ & $0.27$ & \cite{Blake:2012pj} \\
  $0.60$ & $0.390 \pm 0.063$ & $0.27$ & \cite{Blake:2012pj} \\
  $0.73$ & $0.437 \pm 0.072$ & $0.27$ & \cite{Blake:2012pj} \\
  $0.18$ & $0.36 \pm 0.09$ & $0.27$ & \cite{Blake:2013nif} \\
  $0.38$ & $0.44 \pm 0.06$ & $0.27$ & \cite{Blake:2013nif} \\
  $1.4$ & $0.482\pm0.116$ & $0.27$ & \cite{Okumura:2015lvp} \\
  $0.02$ & $0.428_{-0.045}^{+0.048}$ & $0.3$ & \cite{Huterer:2016uyq} \\
  $0.6$ & $0.55 \pm 0.12$ & $0.3$ & \cite{Pezzotta:2016gbo} \\
  $0.86$ & $0.40 \pm 0.11$ & $0.3$ & \cite{Pezzotta:2016gbo} \\
 $0.03$&$0.404^{+0.082}_{-0.081}$ &$0.312$ & \cite{Qin:2019axr} \\
 $0.013$ & $0.46 \pm 0.06$ & $0.315$ & \cite{Avila:2021dqv} \\
 $0.15$ & $0.53 \pm 0.16$ & $0.31$ & \cite{eBOSS:2020yzd}  \\
 $0.38$ & $0.500 \pm 0.047$ & $0.31$ & \cite{eBOSS:2020yzd}  \\
 $0.51$ & $0.455 \pm 0.039$ & $0.31$ & \cite{eBOSS:2020yzd}  \\
 $0.70$ & $0.448 \pm 0.043$ & $0.31$ & \cite{eBOSS:2020yzd}  \\
 $0.85$ & $0.315 \pm 0.095$ & $0.31$ & \cite{eBOSS:2020yzd}  \\
 $1.48$ & $0.462 \pm 0.045$ & $0.31$ & \cite{eBOSS:2020yzd}  \\
\end{tabular}
\end{ruledtabular}
\end{center}
\end{table}

The results from Ref.~\cite{Song:2008qt} are obtained from 2dFGRS data, by extrapolating the results from Ref.~\cite{2004MNRAS.353.1201P}, where bounds at $z=0$ were obtained through a spherical harmonics analysis.
A reconstruction of cosmological large-scale flows in the nearby Universe using the SFI++ Tully-Fisher galaxy sample and the 2MASS redshift survey (2MRS) is presented in Ref.~\cite{Davis:2010sw}, from which the value of $f\sigma_8$ is extracted. In Ref.~\cite{Turnbull:2011ty}, type Ia supernovae (SNeIa) are used as peculiar velocity tracers.
The value of $f\sigma_8$ presented in Ref.~\cite{Blake:2012pj} is obtained from the 2D power spectra measured by the WiggleZ Dark Energy Survey at three overlapping redshift bins. The corresponding covariance matrix has been taken into account in the statistical analysis.
Results for two independent redshift bins are presented in Ref.~\cite{Blake:2013nif}, obtained with the multi-tracer method, from the  Galaxy and Mass Assembly (GAMA) survey.
Using a spectroscopic sample of 2,783 emission-line galaxies from the FastSound survey with the Subaru Telescope \cite{Okumura:2015lvp}, a constraint was derived based on the redshift-space correlation function.
The constraint from Ref.~\cite{Huterer:2016uyq} was derived using a combination of a compilation of low-redshift SNeIa and distances from the fundamental plane of 6dFGS galaxies.
In Ref.~\cite{Pezzotta:2016gbo}, the authors derived their constraint by analyzing the anisotropic galaxy clustering from the final data release of the VIPERS survey, using two different redshift bins.
The constraint from Ref.~\cite{Qin:2019axr} was derived by combining the density and velocity fields from the 2MTF and 6dFGSv surveys.
Using HI line extra-galactic sources from the Arecibo Legacy Fast ALFA (ALFALFA), a constraint on the normalized growth rate parameter is presented in Ref.~\cite{Avila:2021dqv}.
Finally, six measurements of the growth rate parameter from redshift-space distortions (RSD) were presented in the completed SDSS-IV extended Baryon Oscillation Spectroscopic Survey \cite{eBOSS:2020yzd}. These measurements come from the Main Galaxy Sample (MGS), two redshift bins of the BOSS galaxy sample, and three redshift bins from eBOSS, covering Luminous Red Galaxies (LRG), Emission Line Galaxies (ELG), and Quasars.

To obtain the data presented in Table \ref{tab:data-fs8}, redshifts must be converted into distances This conversion is typically performed by the respective collaborations using a fiducial cosmology. For this reason, Table \ref{tab:data-fs8} includes an entry specifying the value of $\Omega_m$ assumed in the fiducial cosmology. \textcolor{black}{
The growth rate combination $f\sigma_8$ is extracted from redshift-space distortions (RSD) in galaxy clustering, where the inferred growth rate $f$ depends on the radial conversion of redshift differences to comoving distances scaling as $s_\parallel \propto 1/H(z)$, while the amplitude $\sigma_8$ reflects the transverse clustering scaling as $s_\perp \propto D_A(z)$. When data are analyzed assuming a fiducial cosmology $\bigl(H^{\rm fid}(z),D_A^{\rm fid}(z)\bigr)$ but compared to theory evaluated in a different cosmology \(\bigl(H(z),D_A(z)\bigr)\), the parameters require separate corrections:
\begin{equation}
f^{\rm fid} \approx f^{\rm true} \times \frac{H(z)}{H^{\rm fid}(z)}, \quad
\sigma_8^{\rm fid} \approx \sigma_8^{\rm true} \times \frac{D_A(z)}{D_A^{\rm fid}(z)}.    
\end{equation}
Therefore, to account for the use of a fiducial cosmology in the data, the growth rate measurements are rescaled using the ratio of $H(z)D_A(z)$ between the tested cosmology and the fiducial model, as given by \cite{Nesseris:2017vor,Macaulay_2013}.:
\begin{equation}
    {\rm ratio}(a,\boldsymbol{\theta}) = \frac{H(a,\boldsymbol{\theta})D_A(a,\boldsymbol{\theta})}{H^{\rm fid}(a)D^{\rm fid}_A(a)}.   
\end{equation}
 }

This formulation ensures to define the negative log likelihood ($\chi^2=-2 \ln \mathcal{L}$)
\begin{equation}\label{likelihood}
    -2 \ln \mathcal{L} (\boldsymbol{\theta}, \sigma_8)=V^i\,C_{ij}^{-1}\,V^j
\end{equation}
where $C_{i,j}$ is the covariance matrix and 

\begin{eqnarray}
    V^i(\boldsymbol{\theta},\sigma_8) = f\sigma_{8,i}-{\rm ratio}(a_i,\boldsymbol{\theta)} f\sigma_8(a_i,\boldsymbol{\theta},\sigma_8).
\end{eqnarray}
Here, $f\sigma_{8,i}$ is the $i$th mean value reported in Table \ref{tab:data-fs8}  while $f\sigma_8(a_i,\boldsymbol{\theta}, \sigma_8)$ is its theoretical prediction, for cosmological parameters ($\boldsymbol{\theta}, \sigma_8)$. The term ${\rm ratio}(a_i,\boldsymbol{\theta})$ rescales the result to the fiducial cosmology, to make the comparison on an equal footing.
Apart from the correlated data from the overlapping redshift bins of the WiggleZ dataset, where correlations are taken into account, the remaining datasets are independent. Therefore, for these datasets, we assume uncorrelated data with $C_{i,i}=\sigma_i^2$ where $\sigma_i$ corresponds to the observational error in Table \ref{tab:data-fs8} and $C_{i,j}=0$ for $i \neq j$. The block of the covariance matrix that corresponds to the WiggleZ data is taken from Ref.~\cite{Blake:2012pj}.
%
%
%
%
%

\section{Results}\label{results}

In this section, we present the results from the statistical analysis performed using the solutions obtained with the PINN bundle method and the data described in Sec.~\ref{data}. %

\begin{table}[h]
\centering
\caption{Flat priors for all the parameters considered in this work.}
\begin{ruledtabular}
\begin{tabular}{cccc}
- & $\Omega_m$ & $\sigma_8$ & $g_a$ \\
\hline
$\Lambda$CDM & [0.05, 0.7] & [0.5, 1.5] & - \\
\hline
MG & [0.05, 0.7] & [0.5, 1.5] & [-3$\times 10^{-2}$, 3$\times 10^{-2}$] \\
\end{tabular}
\end{ruledtabular}
\label{priors}
\end{table}

\begin{figure}[h!]
    \centering
    \includegraphics[width=\linewidth]{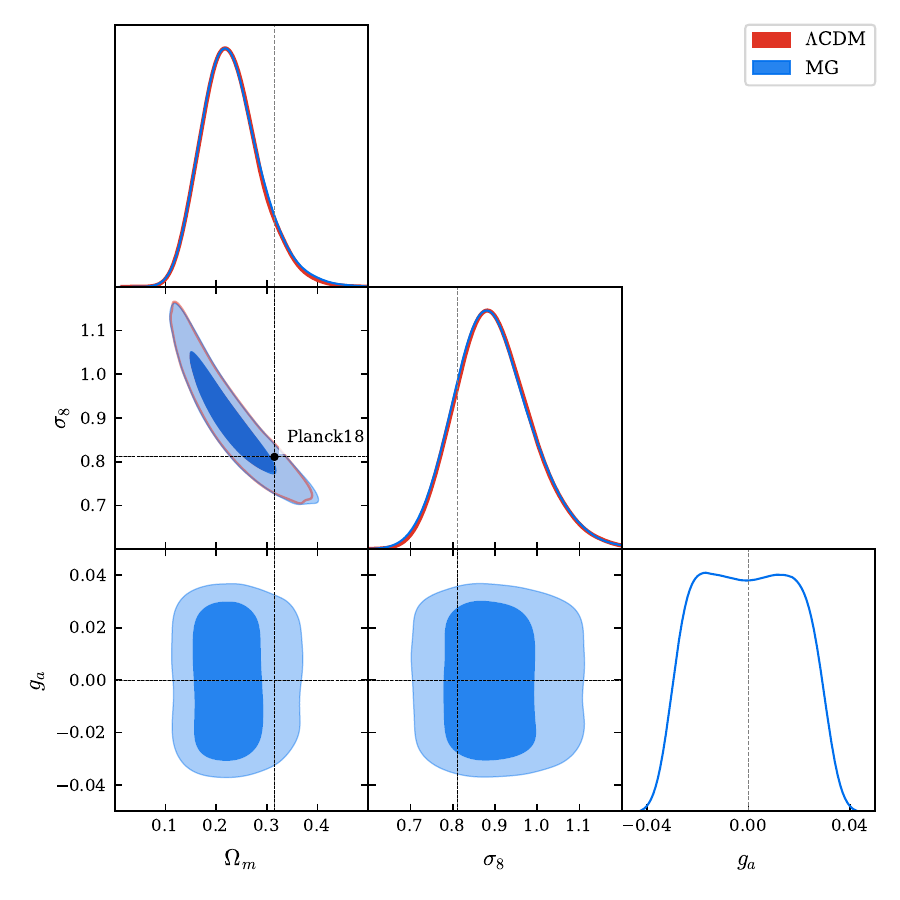}
    \caption{Results of the statistical analysis for the $\Lambda$CDM (red) and phenomenological gravity (blue) model. The darker and brighter regions correspond to the
68\% and 95\% confidence regions respectively. The plots on
the diagonal show the posterior probability density for each of the
free parameters of the model. The black dot and dotted lines correspond to the mean value of $\Omega_m$ and $\sigma_8$ estimated by Planck18 assuming $\Lambda$CDM. }
    \label{cornerplots}
\end{figure}

We used the Markov chain Monte Carlo (MCMC) method to sample the posterior distributions, with the negative log likelihood function defined in Eq.~\eqref{likelihood} and the priors defined in Table \ref{priors}. The priors on $g_a$ are informative as was discussed in Sec.~\ref{theory}. The 68\% and 95\% confidence
level contours of the posterior probability distribution of
the parameters,  are shown in Fig.~\ref{cornerplots} for $\Lambda$CDM and the phenomenological modified gravity model.
Table \ref{results} presents the 68\% and 95\% confidence intervals, as well as the mean and best-fit values for the parameters of both models.

\begin{figure}[h!]
    \centering
    \includegraphics[width=\linewidth]{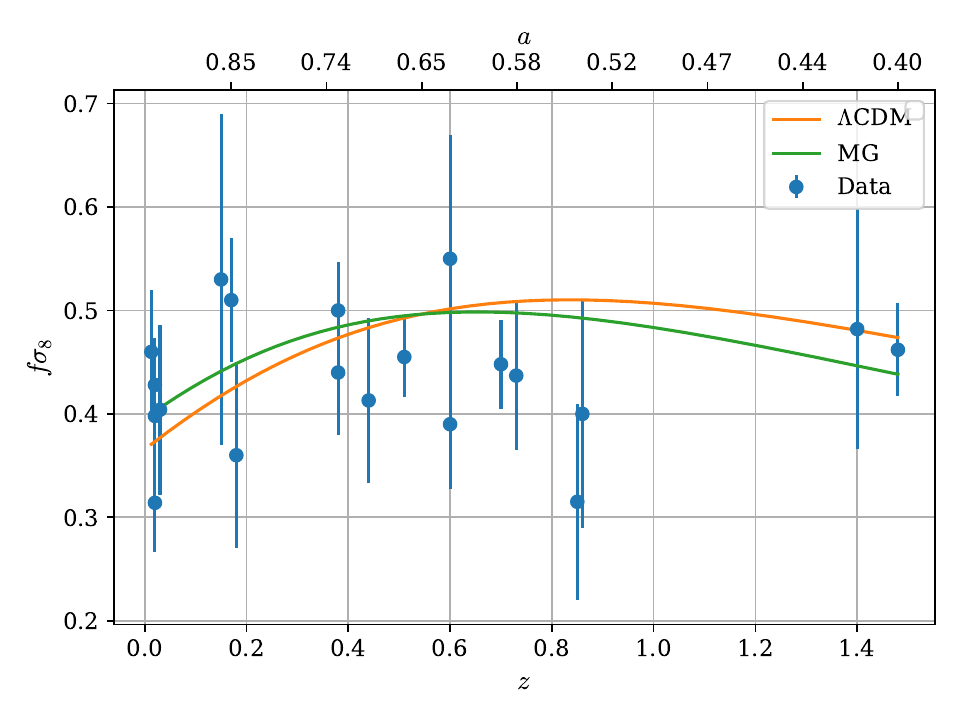}
    \caption{$f\sigma_8$ vs $z$. In the plot we show the best fit for $\Lambda$CDM (orange) and the phenomenological modified gravity model (green) together with the data. }
    \label{compLCDM-MG-data}
\end{figure}

\textcolor{black}{Our results (see Fig.\ref{cornerplots} and Table \ref{Tab: table_1}) show almost no differences between the 2-dimensional contour plots and 1-dimensional posteriors of the modified gravity model and those of $\Lambda$CDM. Consequently, the inferred values of $\Omega_m$ and $\sigma_8$ shown in Table \ref{Tab: table_1} are the same for both models. The reason for this lies in the restrictive informative prior that we have assumed for $g_a$ (see Table \ref{priors} and the discussion in Section \ref{model_modgrav}). This prior follows from limits in the variation in $G$, established through comparisons between the theoretical predictions of solar models and the solar p-model frequency observations. For the values of $g_a$ allowed by the mentioned, the predictions of the modified gravity model show no major differences from the ones of $\Lambda$CDM. In turn, this is the reason why it is not possible to further constrain $g_a$.}
In addition, Fig.~\ref{cornerplots} shows no degeneracy between $g_a$ and the other cosmological parameters, which is also a result of the restrictive prior on $g_a$.  When a less restrictive prior is allowed, degenerations appear in the planes $g_a-\Omega_m$ and $g_a -\sigma_8$, as expected from Eqs.\eqref{XY}.
Figure \ref{compLCDM-MG-data} shows the evolution of $f\sigma_8(z)$ for the best-fit models of $\Lambda$CDM and modified gravity together, along with the data used in this paper.

\begin{figure}[h!]
    \centering
    \includegraphics[width=\linewidth]{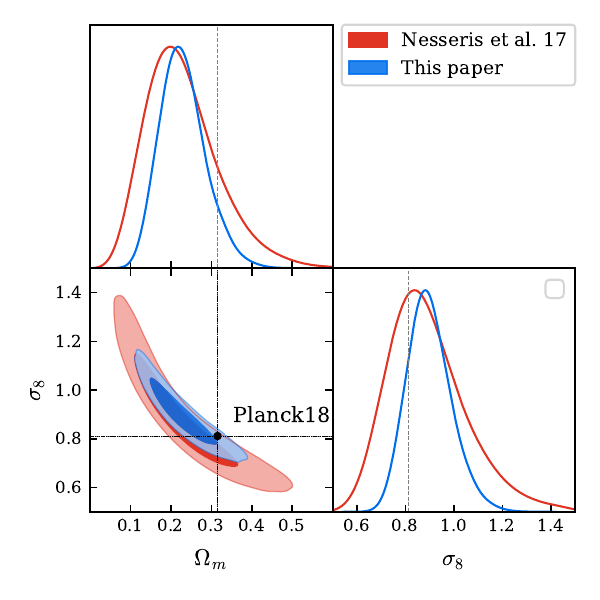}
    \caption{Results of the statistical analysis using the data set compiled by Ref.~\cite{Nesseris:2017vor} (red) and the  data set described in Table \ref{data} (blue).  \textcolor{black}{The latter includes updated data from the GAMA,VIPERS, 6dFGS, ALFALFA surveys as well as  SDSS-IV  RSD measurements, none of which were  considered in Ref.~\cite{Nesseris:2017vor} }. The black dot and dotted lines correspond to the mean value of $\Omega_m$ and $\sigma_8$ estimated by Planck18 assuming $\Lambda$CDM. 
}
    \label{comp_old_new_data}
\end{figure}

The data set compiled in Table \ref{data} enables us to obtain more stringent constraints on $\Omega_m$ and $\sigma_8$ than the ones obtained in previous works \cite{Nesseris:2017vor,2024arXiv241105965Z}. This can be seen in Fig. \ref{comp_old_new_data}, where the 2-dimensional contours and the 1-dimensional posteriors of the $\Lambda$CDM model are shown for both data sets (the one described in Table \ref{data} and the one used in Refs \cite{Nesseris:2017vor,2024arXiv241105965Z}). However, the inferred 95\% confidence intervals of $\Omega_m$ and $\sigma_8$  are consistent with those found in earlier works. Furthermore, the degeneracy shown in the $\Omega_m-\sigma_8$ is similar to what has been reported in these works. Furthermore, the mean values of $\Omega_m$ and $\sigma_8$ inferred by the Planck collaboration \cite{Aghanim:2018eyx} using CMB data and assuming the $\Lambda$CDM model are within the 95\% confidence contours of the $\Omega_m-\sigma_8$ plane inferred in this work.

\begin{table}[h!]
    \centering
    \caption{Constraints on the free parameters for $\rm{\Lambda CDM}$ and the phenomenological modified gravity model.}
    \begin{ruledtabular}
    \begin{tabular}{cccccc}
        Model &  & $\Omega_m$ & $\sigma_8$ & $g_a$ & $\chi^2_{\rm red}$ \\
        \midrule
        \multirow{4}{*}{$\Lambda$CDM} & 68$\%$ & [0.20, 0.25] & [0.85, 0.93] & & \\
        & 95$\%$ & [0.15, 0.33] & [0.76, 1.06] & & \\
        & Mean & 0.23 & 0.90 &  &  \\
        & Best fit & 0.17 & 0.98 & & 1.06 \\
        \midrule
        \multirow{4}{*}{MG} & 68$\%$ & [0.20, 0.25] &  [0.85, 0.93] & [-0.01, 0.01] &  \\
        & 95$\%$ & [0.15, 0.33] & [0.76, 1.06] & [-0.03, 0.03] & \\
        & Mean & 0.23 & 0.90 & -3$\times 10^{-4}$ \\
        & Best fit & 0.22 & 0.92 & 0.02 & 0.95 \\
    \end{tabular}
    \end{ruledtabular}
    \label{Tab: table_1}
\end{table}


\section{Summary and Conclusions} \label{conclusions}

We obtained the evolution of the linear matter perturbations for both the $\Lambda$CDM model and a phenomenological modified gravity model, using the PINN bundle method. The PINN bundle method has previously been applied to solve the equations governing the background dynamics of the Universe \cite{our_previous_paper,2024PhRvD.109l3514C}. In this paper, we show the application to a different scenario, namely, the growth of matter perturbations. A key novelty of this work is the calculation of an error bound to approximate the theoretical prediction of the observable $f\sigma_8$, using only the PINN-based solution and its residuals. In contrast, in our previous work, we assessed the accuracy of the PINN bundle method, by comparing its outcomes with numerical solutions. This represents a step forward for the PINN bundle method in providing a rule to calculate its own error. We also highlight the use of an updated dataset for the observable $f\sigma_8$, which provided more stringent constraints on the cosmological parameters $\Omega_m$ and $\sigma_8$ and demonstrated consistency with the latest CMB data.

The modified gravity model considered here is a phenomenological model with a $\Lambda$CDM background and a varying Newton's constant, motivated by modified gravity models whose variation is constrained to meet current limits. 
This work marks the first step in solving the matter perturbation equation for a modified gravity model derived from first principles. 
In that case, we will face  additional complexity due to the dependence of the equation and its initial condition with the Fourier mode. Therefore, we expect that the PINN bundle method will provide a  substantial reduction of the computational times of the inference process. 



\begin{acknowledgments}
C.G.S. would like to thank Domenico Sapone for helpful discussions on the data sets. The computations performed in this paper, which used GPUs, were run on the FASRC Cannon cluster supported by the FAS Division of Science Research Computing Group at Harvard University.  S.L. is supported by grant PIP 11220200100729CO CONICET and grant 20020170100129BA UBACYT.

\end{acknowledgments}

\appendix
\section{Error bound on \texorpdfstring{$X$}{X}}\label{boundX}

In this appendix we derive an expression for the error in the variable $X$ and its corresponding bound. We recall that  

\begin{equation}
    Y \equiv \frac{dX}{dN}\equiv \tilde{Y}+\eta^Y=\frac{d\tilde{X}}{dN}+\frac{d\eta^X}{dN},
\end{equation}then

\begin{equation}
    \frac{d\eta^X}{dN}=-\mathcal{R}_X+\eta^Y,
\end{equation}where $\mathcal{R}_X(N)$ is the residual of Eq.~\eqref{ec2}. Therefore we can integrate this equation to get

\begin{equation}
    \eta^X(N)=\int_{-1}^N[-\mathcal{R}_X(N^\prime)+\eta^Y(N^\prime)]dN^\prime,
\end{equation}where $N_{\rm ini} = -1$ was used. Now we take the absolute value and use Eq.~\eqref{defetai} to write


\begin{equation}\label{etax2}
    |\eta^X(N)| = \left|-\int_{-1}^N\mathcal{R}_X(N^\prime)dN^\prime + \sum_{j=0}^{+\infty}\int_{-1}^N\eta_j^Y(N^\prime)dN^\prime\right|.
\end{equation}
Now we define the truncation order $J_X$ as follows

\begin{equation}
    \sum_{j=0}^{+\infty}\eta_j^Y = \sum_{j=0}^{J_X}\eta_j^Y+\sum_{j=J_X+1}^{+\infty}\eta_j^Y.
\end{equation}
Using this last definition in Eq.~\eqref{etax2} we get

\begin{equation}
\begin{split}
    |\eta^X(N)|\leq&\left|-\int_{-1}^N\mathcal{R}_X(N^\prime)dN^\prime + \sum_{j=0}^{J_X}\int_{-1}^N\eta_j^Y(N^\prime)dN^\prime\right| \\
    &+\DOWNCBRACE{\sum_{j=J_X+1}^{+\infty}\int_{-1}^N|\eta_j^Y|dN^\prime}{E:=}
\end{split}
\end{equation}

Now we can use the fact that $|\eta_j^Y(N)|\leq R[R(N+1)]^j$ (the proof can be founded in Appendix B of Ref.~\cite{2024arXiv241113848C}) to bound $E$ 

\begin{eqnarray}
    E&\leq& \sum_{j=J_X+1}^{+\infty}R^{j+1}\int_{-1}^N (N^\prime+1)^jdN^\prime \nonumber\\
    &\leq &\sum_{j=J_X+1}^{+\infty}R^{j+1} \frac{(N+1)^{j+1}}{j+1} \nonumber\\
    &\leq &[R(N+1)]^{J_X+1}\sum_{j=J_X+1}^{+\infty} \frac{[R(N+1)]^{j-J_X}}{j+1}
    \nonumber \\
    &\leq & [R(N+1)]^{J_X+1}\sum_{j=1}^{+\infty} \frac{[R(N+1)]^{j}}{j+J_x+1}\nonumber\\
    &< &[R(N+1)]^{J_X+1}\sum_{j=1}^{+\infty}\frac{[R(N+1)]^j}{j},
\end{eqnarray}
where we used that if $J_x+1>0$, then $\frac{[R(N+1)]^{j}}{j+J_x+1}< \frac{[R(N+1)]^{j}}{j}$. Also, note that this sum converges $\forall N\in [-1,0]$ only if $R<1$, so this condition imposes constraints on the residual of the neural networks. Now we use that $\sum_{j=1}^{+\infty}\frac{x^j}{j}=-\ln{(1-x)}$ with $x = R(N+1)$ and define $\mathcal{B}_X$ to bound the value of $|\eta^X|$, so

\begin{equation}\label{BX}
\begin{split}
    \mathcal{B}_X(N;J_X):=\left|-\int_{-1}^N\mathcal{R}_X(N^\prime)dN^\prime+\sum_{j=0}^{J_X}\int_{-1}^N\eta^Y_j(N^\prime)dN^\prime\right| \\
    -\left[R(N+1)\right]^{J_X+1}\ln{[1-R(N+1)]}.
\end{split}
\end{equation}


The integrals required for computing both $\mathcal{B}_Y$ and $\mathcal{B}_X$ in this work  were evaluated  using the trapezoidal rule with $10^{4}$ integration points.

\bibliography{biblio}

\end{document}